\documentclass[journal]{IEEEtran}

\usepackage{algorithm}
\usepackage{algorithmicx}
\usepackage{algpseudocode}
\usepackage{amsmath}
\usepackage{graphicx}
\usepackage{overpic}
\usepackage{multicol}
\usepackage{multirow}
\usepackage{cite}
\usepackage{array}
\usepackage{booktabs}
\usepackage{mathrsfs}
\usepackage{amsfonts}
\usepackage{color}
\usepackage{float}
\usepackage{fancyhdr}
\usepackage{threeparttable}
\usepackage{ragged2e}
\usepackage{amssymb}

\renewcommand{\raggedright}{\leftskip=0pt \rightskip=0pt plus 0cm}
\renewcommand{\arraystretch}{1.5}

\begin{document}
\title{SS-CTML: Self-Supervised Cross-Task Mutual Learning for CT Image Reconstruction}
\author{Gaofeng Chen, Yaoduo Zhang, Li Huang, Pengfei Wang, Wenyu Zhang, Dong Zeng, \\
Jianhua Ma, and Ji He
\thanks{This work was supported in part by the National Natural Science Foundation of China under Grant Nos. 62001208, U21A6005, U1708261. \it{(Corresponding author: Ji He.)}}
\thanks{Gaofeng Chen is with School of Biomedical Engineering, Guangzhou Medical University, Guangzhou 511436, China, and also with Sun Yat-sen Memorial Hospital, Sun Yat-sen University, Guangzhou 510120, China.}
\thanks{D.~Zeng is with School of Biomedical Engineering, Southern Medical University, Guangzhou 510515, China.}
\thanks{J.~Ma is with School of Life Science and Technology, Xi'an Jiaotong University, Xi'an 710049, China.}
\thanks{Y.~Zhang, L.~Huang, P.~Wang, W.~Zhang, and J.~He are with School of Biomedical Engineering, Guangzhou Medical University, Guangzhou 511436, China, and also with the Fourth Affiliated Hospital, Guangzhou Medical University, Guangzhou 511300, China (e-mail: heji@gzhmu.edu.cn).}
}

\maketitle

\begin{abstract}
Supervised deep-learning (SDL) techniques with paired training datasets have been widely studied for X-ray computed tomography (CT) image reconstruction. However, due to the difficulties of obtaining paired training datasets in clinical routine, the SDL methods are still away from common uses in clinical practices. In recent years, self-supervised deep-learning (SSDL) techniques have shown great potential for the studies of CT image reconstruction. In this work, we propose a self-supervised cross-task mutual learning (SS-CTML) framework for CT image reconstruction. Specifically, a sparse-view scanned and a limited-view scanned sinogram data are first extracted from a full-view scanned sinogram data, which results in three individual reconstruction tasks, i.e., the full-view CT (FVCT) reconstruction, the sparse-view CT (SVCT) reconstruction, and limited-view CT (LVCT) reconstruction. Then, three neural networks are constructed for the three reconstruction tasks. Considering that the ultimate goals of the three tasks are all to reconstruct high-quality CT images, we therefore construct a set of cross-task mutual learning objectives for the three tasks, in which way, the three neural networks can be self-supervised optimized by learning from each other. Clinical datasets are adopted to evaluate the effectiveness of the proposed framework. Experimental results demonstrate that the SS-CTML framework can obtain promising CT image reconstruction performance in terms of both quantitative and qualitative measurements.
\end{abstract}


\section{Introduction}
\label{sec:introduction}
\IEEEPARstart{X}{-ray} computed tomography (CT) is widely applied in clinical diagnosis and radiation treatment, since it can provide precise anatomy information for patients. However, excessive exposure radiation dose in CT examinations is harmful to patients, which has becoming a concerning problem\cite{brenner2007computed}. Therefore, low-dose CT (LDCT) imaging techniques have become a popular study to reduce patient damages from X-ray radiation. The most common strategies of LDCT include sparse-view CT (SVCT), limited-view CT (LVCT), and low tube-current/mAs with full-view CT (low-mAs FVCT) scannings, all of which will inevitably lead to low quality of CT images reconstructed by filtered back-projection (FBP) method. In order to solve this problem, advanced reconstruction algorithms are highly desirable. In the past decade, supervised deep-learning (SDL) methods have been widely studied for CT image reconstruction and have achieved great improvements\cite{kang2018deep,jin2017deep,chen2022sam}. However, clinical paired training datasets are usually impractical to acquire, due to the patient unavoidable respiratory movement and/or the irrational second radiation exposure.

In recent years, self-supervised deep-learning (SSDL) techniques have shown great potential for CT image reconstruction, which usually require only the unlabeled degraded data. Generally, these strategies can be divided into four categories. The first strategy is model-based deep learning techniques, which usually incorporate neural network reconstruction architectures with a set of model-based objective functions so as to reduce noise-induced artifacts in CT images iteratively\cite{meng2020semi,li2022noise,zang2021intratomo}. However, designing such model-based objective functions usually requires elaborate prior assumptions in the imaging process, which is not an easy task.

The second strategy is generally with the form of `which2which', such as Noise2Noise\cite{lehtinen2018noise2noise}, Neighbor2Neighbor\cite{huang2021neighbor2neighbor}, and Noise2Self\cite{batson2019noise2self}, etc. These methods were initially developed for natural image denoising, and then rapidly expanded for CT imaging. The main idea is first generating a surrogate paired dataset with only the unlabeled LDCT data in sinogram and/or image domains, and then using it to train a model for sinogram and/or image restoration\cite{hendriksen2020noise2inverse,lagerwerf2020noise2filter,li2020sacnn,wu2023unsharp}. For example, Hendriksen \textit{et al.} proposed a noise2inverse method for low-mAs FVCT image denoising by constructing a pseudo paired training dataset in sinogram domain \cite{hendriksen2020noise2inverse}. Wu \textit{et al.} proposed a neighbor2neighbor method to generate paired training data and introduced unsharp guided filtering for self-supervised low-mAs FVCT image denoising\cite{wu2023unsharp}. Although these methods have achieved promising performance, the over-simplified noise assumptions (such as, zero-mean Gaussian, and/or independent-identical distributions) in this kind of methods are not always suitable for LDCT imaging, because the noise distributions of which are much more complicated in real clinical practices.

The third strategy is implicit neural representation (INR)-based techniques. With the help of image continuity prior imposed by the multi-layer perceptron (MLP), a lot of INR-based methods have been developed for SVCT or LVCT image reconstruction\cite{wu2023self,zang2021intratomo,lee2024iterative}. Although the performance of these methods for SVCT/LVCT image reconstruction are promising, the training process of the INR-based methods is computationally complicated, which is not suitable for real clinal practices.

Recently, diffusion model-based techniques have been widely studied and can also be regarded as another kind of self-supervised learning strategy. Diffusion models are a class of deep generative model, which are trained to model the process of Markov transition from a simple random distribution to complicated data distribution so that the target sample can be generated through sequential stochastic transitions\cite{liu2023dolce,ho2020denoising}. Apart from natural image generation and image denoising/restoration\cite{xie2023diffusion,kawar2022denoising}, diffusion models have also gained widespread attention in CT imaging, such as, low-mAs FVCT image reconstruction\cite{gao2022cocodiff,liu2023diffusion}, SVCT image reconstruction\cite{guan2023generative,yang2023dual} and LVCT image reconstruction\cite{liu2023dolce,chung2023solving}. However, similar to the INR-based methods, the practical applications of diffusion models are still limited, due to the large computational occupation and long inference time. Such concerns should be first resolved before both kind of methods can be adopted for real clinical use.

Although lots of self-supervised learning methods have been proposed for CT image reconstruction, most of them just focus on handling one single image reconstruction task (e.g., either low-mAs FVCT, SVCT, or LVCT image reconstruction), where the correlations among different reconstruction tasks are ignored. It is noted that the SVCT and LVCT image reconstruction tasks are hidden in low-mAs FVCT reconstruction task, because the SVCT or LVCT scanned sinogram data can be easily extracted from the low-mAs FVCT scanned sinogram data. Considering the ultimate goals of these three tasks are all to reconstruct high-quality CT images from the same object being scanned, we claim that the performance of different reconstruction tasks can be simultaneously improved by learning from each other without the high-quality normal-dose FVCT sinogram and/or images. Inspired by the above observations, in this work, we propose a novel self-supervised cross-task mutual learning framework for CT image reconstruction. For simplicity, the proposed framework is denoted as SS-CTML. The main contributions of this work can be summarized as follows:

(1) We propose a novel self-supervised cross-task mutual learning framework, which fully considers the correlations among different reconstruction tasks for CT image reconstruction. With a set of cross-task mutual learning objective functions, the proposed framework can be self-supervised optimized to improve the performance of the low-mAs FVCT reconstruction, SVCT reconstruction, and LVCT reconstruction at the same time.

(2) We establish an effective network architecture for the SS-CTML framework, which consists of three dual-domain (i.e., sinogram and image domains) subnetworks. In each subnetwork, we first construct a prior neural module, where the FBP reconstruction (i.e., either the low-mAs FVCT or the SVCT/LVCT reconstruction) is implemented as the network input. Then, the output image is considered as a prior, which is further forward-projected into the sinogram domain to compensate LDCT sinogram data. Finally, the compensated sinogram data is fed into a dual-domain neural module for image reconstruction.

(3) Experimental results demonstrate that the proposed SS-CTML framework can achieve promising performances for low-mAs FVCT, SVCT, and LVCT image reconstruction in terms of noise suppression, artifact reduction and fine structure preservation.
\section{Related Work}
\subsection{Mutual Learning}
Mutual learning is a specific form of knowledge distillation (KD), which is first proposed by Hinton \textit{et al.}\cite{hinton2015distilling}, and is an effective technique to transfer the learned knowledge from a powerful and large network (a.k.a. teacher network) to a weak and small network (a.k.a. student network). In recent years, KD has been widely applied for various medical image processing tasks, including medical image segmentation\cite{qin2021efficient} and medical CT imaging\cite{wang2023self}. For example, Wang \textit{et al.} proposed a self-supervised guided KD (SGKD) framework for LDCT image restoration, which enables the guidance of supervised learning using the results generated by self-supervised learning\cite{wang2023self}. On the contrary, mutual learning enables several student networks to learn from each other without a teacher network\cite{zhang2018deep}. Fang \textit{et al.} proposed a cross-view mutual distillation learning method for medical CT image synthesis, which improves the inter-slice resolution of CT volumetric image by constructing three student networks for coronal, axial and sagittal view of the CT volumetric image\cite{fang2022incremental}. Different from the above study that applied coronal, axial and sagittal mutual learning for one single CT image synthesis task, this paper proposes an FVCT, SVCT, and LVCT mutual learning framework that can simultaneously perform different CT image reconstruction tasks, i.e., the low-mAs FVCT reconstruction, the SVCT reconstruction, and the LVCT reconstruction.
\begin{figure*}[htb]
\centerline{\includegraphics[scale=0.86]{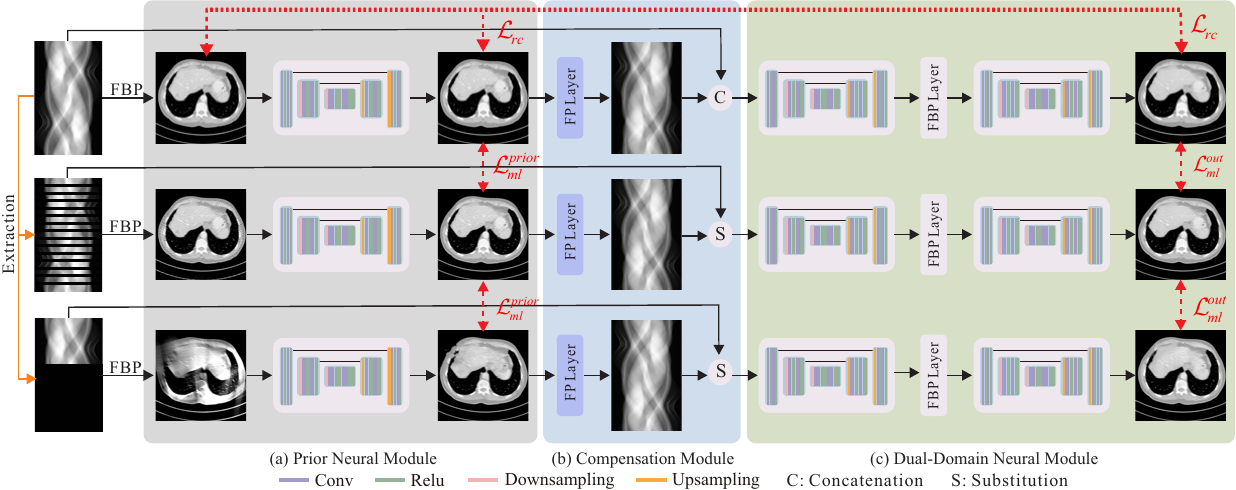}}
\caption{Overview of the proposed SS-CTML framework. The mutual learning of cross-task (i.e., low-mAs FVCT, SVCT, and LVCT reconstruction tasks) is utilized to regularize the training process. The proposed network architecture mainly consists of prior neural modules(PNMs), compensation modules and dual-domain neural modules(DDNMs).}
\label{fig1}
\end{figure*}
\subsection{Dual-Domain Network Architecture}
Dual-domain network architecture has been widely studied, which can constrain measurement consistency in sinogram domain and reduce pixel-wise error in image domain, and has shown superior performance in low-mAs FVCT, SVCT and LVCT reconstructions\cite{yang2022low, wu2021drone, chen2022sam}. For example, He \textit{et al.} proposed iRadonMap with a sinusoidal back-projection module to enable dual-domain (i.e., sinogram and image domains) learning for inverse Radon transform approximation\cite{8950464}, and then extended it to the downsampled imaging geometric modeling technique for low-mAs FVCT and SVCT reconstructions\cite{9410296}. Yang \textit{et al.} proposed sinogram inner-structure transformer (SIST) to perform projection restoration in sinogram domain and utilized image reconstruction module to remove noise in image domain\cite{yang2022low}. Wu \textit{et al.} extended the sparse-view data to full-view one through a U-net architecture in sinogram domain and then adopted a Wasserstein generative adversarial network (WGAN) to remove the SVCT artifacts in image domain\cite{wu2021drone}. Chen \textit{et al.} proposed a self-augmented multi-stage dual-domain neural network (Sam's Net) for LVCT reconstruction, where a substitution module was implemented as a online self-constrained operation to combine information from both the output projection in previous stage and the input projection at current stage\cite{chen2022sam}. Different from the most existing dual-domain network architectures, which are designed for supervised learning framework, we construct a prior neural module as the self-constraint for the dual-domain learning without multi-stage iterations, which is specifically designed for the self-supervised mutual optimization framework.
\section{Method}
The main goal of this study is to develop a self-supervised learning framework for CT image reconstruction with only one LDCT training dataset, i.e., $S_{ld}=\{p_{ld}^{i}, \mu_{ld}^{i}\}_{i=1}^{M}$, which contains the low-mAs FVCT sinogram data $p_{ld}$ and the corresponding FBP reconstruction $\mu_{ld}$. Here, $ld$ denotes for low-dose, $M$ denotes the number of training samples in the dataset. In order to achieve this goal, we explore the intrinsic correlations among different reconstruction tasks (i.e., low-mAs FVCT image reconstruction, SVCT image reconstruction, and LVCT image reconstruction) within the LDCT training dataset $S_{ld}$, and propose a novel self-supervised cross-task mutual learning framework for CT image reconstruction, denoted as SS-CTML. In the following, we first present an overview of the proposed SS-CTML framework in Section \ref{section:A}. Then, we describe the network architecture that specific designed for cross-task mutual optimization in Section \ref{section:B}, followed by details of cross-task mutual learning objective functions in Section \ref{section:C}.
\begin{figure}[htb!]
\centerline{\includegraphics[width=\columnwidth]{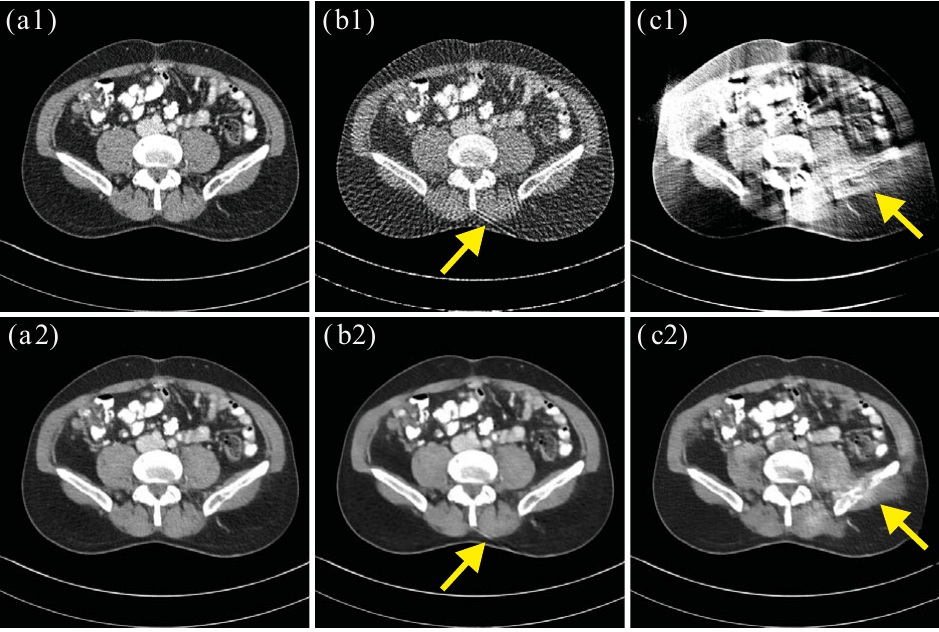}}
\caption{The comparison of different reconstruction tasks. From (a) to (c): low-mAs FVCT, SVCT and LVCT reconstruction tasks, respectively. The first row and second row are the reconstruction results of FBP and cross-task mutual learning in image domain, respectively. The display window is [-120, 280] HU.}
\label{fig2}
\end{figure}
\subsection{Framework Overview}
\label{section:A}
The proposed SS-CTML framework is illustrated in Fig. \ref{fig1}. The general idea of the proposed framework is to construct three different CT reconstruction tasks (i.e., low-mAs FVCT image reconstruction, SVCT image reconstruction, and LVCT image reconstruction) from the LDCT training dataset $S_{ld}$, and then constrain these three reconstruction tasks to learn from each other. In data preparation phase, we first construct another two datasets (i.e., an SVCT training dataset and an LVCT training dataset) from the LDCT training dataset $S_{ld}$, by extracting SVCT sinogram data $p_{sv}$ and LVCT sinogram data $p_{lv}$ from the low-mAs FVCT sinogram data $p_{ld}$, according to two specific SVCT and LVCT imaging geometries, respectively. Here, $sv$ denotes for sparse-view and $lv$ denotes for limited-view. The corresponding SVCT image $\mu_{sv}$ and LVCT image $\mu_{lv}$ are also reconstructed from $p_{sv}$ and $p_{lv}$ via FBP algorithm, respectively. The constructed SVCT training dataset and LVCT training dataset are denoted as $S_{sv}=\{p_{sv}^{i}, \mu_{sv}^{i}\}_{i=1}^{M}$ and $S_{lv}=\{p_{lv}^{i}, \mu_{lv}^{i}\}_{i=1}^{M}$, respectively. The first row of Fig. \ref{fig2} shows three FBP reconstruction results on same anatomic slice from low-mAs FVCT, SVCT, and LVCT sinogram data. As can be seen, the anatomic information in the low-mAs FVCT, SVCT, and LVCT images is contaminated by noise and/or artifacts in different degrees due to different imaging geometries. Nevertheless, the reconstruction information with different imaging geometries can be mutual compensated. Therefore, we claim that the three reconstruction tasks can be mutually optimized by learning from each other.

With the three training datasets, i.e., $S_{ld}$, $S_{sv}$, and $S_{lv}$, we then construct three subnetworks for their corresponding reconstruction tasks, i.e., $f_{ld}$ for low-mAs FVCT reconstruction, $f_{sv}$ for SVCT reconstruction, and $f_{lv}$ for LVCT reconstruction, as shown in Fig. \ref{fig1}. Each subnetwork takes both the sinogram data and FBP image as inputs, then outputs the final reconstruction. The process can be depicted as follows:
\begin{equation}
    \begin{cases}
     \mu_{ld}^{out}=f_{ld}(p_{ld},\mu_{ld}~|~\Omega_{ld})
    \\
    \mu_{sv}^{out}=f_{sv}(p_{sv},\mu_{sv}~|~\Omega_{sv})
     \\
    \mu_{lv}^{out}=f_{lv}(p_{lv},\mu_{lv}~|~\Omega_{lv})
    \end{cases},
\label{eq1}
\end{equation}
where $\mu_{ld}^{out}$, $\mu_{sv}^{out}$, and $\mu_{lv}^{out}$ are the final outputs of the low-mAs FVCT reconstruction subnetwork $f_{ld}$, the SVCT reconstruction subnetwork $f_{sv}$, and the LVCT reconstruction subnetwork $f_{lv}$, respectively. $\Omega_{ld}$, $\Omega_{sv}$, and $\Omega_{lv}$ are the network parameters. It is noted that, for same anatomic slice, $\mu_{ld}^{o}$, $\mu_{sv}^{o}$, and $\mu_{lv}^{o}$ represent the same anatomic information. In another word, the three outputs should be as close as possible, which motivates us to constrain the three reconstruction tasks by letting their correlated three networks to learn from each other. In the following sections, we will present more details about the network architecture and training objective functions in the proposed SS-CTML framework.

\subsection{Network Architecture}
\label{section:B}
Intuitively, the three subnetworks, i.e., $f_{ld}$, $f_{sv}$, and $f_{lv}$ can be simply instantiated as three restoration networks with common network architectures, such as, ResNet or U-net, in which way, the cross-task mutual learning can be simply implemented in image domain. However, we empirically found that the reconstruction performance is not satisfactory enough due the weak fitting ability of such network architecture. The second row of Fig. \ref{fig2} shows the low-mAs FVCT, SVCT, and LVCT restoration results of the cross-task mutual learning that directly applied in the image domain. As can be seen, although such strategies can wipe out majority artifacts and restore anatomic structures for the low-mAs FVCT task, significant artifacts and anatomy distortions are still existing in SVCT and LVCT reconstructions, as indicated by the yellow arrows in Fig. \ref{fig2}. Nevertheless, the three reconstructions can still be fairly good prior images, which should be further utilized.

Therefore, in $f_{ld}$, $f_{sv}$, and $f_{lv}$, we first construct three prior neural modules (PNMs), which are denoted as $f_{ld}^{prior}$, $f_{sv}^{prior}$, and $f_{lv}^{prior}$, respectively. The PNMs take the three contaminated FBP images (i.e., $\mu_{ld}$, $\mu_{sv}$, and $\mu_{lv}$) as inputs and output three prior images (i.e., $\mu_{ld}^{prior}$, $\mu_{sv}^{prior}$, and $\mu_{lv}^{prior}$), which can be represented as:
\begin{equation}
    \begin{cases}
     \mu_{ld}^{prior}=f_{ld}^{prior}(\mu_{ld}~|~\beta_{ld})
    \\
    \mu_{sv}^{prior}=f_{sv}^{prior}(\mu_{sv}~|~\beta_{sv})
     \\
    \mu_{lv}^{prior}=f_{lv}^{prior}(\mu_{lv}~|~\beta_{lv})
    \end{cases},
\label{eq2}
\end{equation}
where $\beta_{ld}$, $\beta_{sv}$, and $\beta_{lv}$ are the learnable parameters. Then, the prior images are forward projected to compensate the contaminated sinogram data (i.e., $p_{ld}$, $p_{sv}$, and $p_{lv}$), as written below:
\begin{equation}
    \begin{cases}
    \tilde{p}_{ld} = A\mu_{ld}^{prior}\oplus p_{ld}
    \\
    \tilde{p}_{sv} = A\mu_{sv}^{prior}\times M_{sv} + p_{sv}\times (1-M_{sv})
     \\
    \tilde{p}_{lv} = A\mu_{lv}^{prior}\times M_{lv} + p_{lv}\times (1-M_{lv})
    \end{cases},
\label{eq3}
\end{equation}
where $A$ is a back-propagative forward projection operator (i.e., FPLayer in Fig. \ref{fig1}); $\oplus$ is the concatenation operator for low-mAs FVCT imaging; $M_{sv}\in[0,1]$ is a binary mask matrix for SVCT imaging, and $1$ denotes the missing information in $p_{sv}$; the definition of $M_{lv}$ for LVCT imaging is similar to that of $M_{sv}$. $\tilde{p}_{ld}$, $\tilde{p}_{sv}$, and $\tilde{p}_{lv}$ are the compensated sinogram data.

With the compensated sinogram data, we further construct three dual-domain neural modules (DDNMs) for the three tasks, which are denoted as $f_{ld}^{dual}$, $f_{sv}^{dual}$, and $f_{lv}^{dual}$, respectively. Specifically, each DDNM contains one sinogram network and one image network, with a back-propagative FBP operator (i.e., FBPLayer in Fig. \ref{fig1}) to bridge them. Therefore, the DDNMs are able to take the three compensated sinogram data (i.e., $\tilde{p}_{ld}$, $\tilde{p}_{sv}$, and $\tilde{p}_{lv}$) as inputs and output the three final reconstructions (i.e., $\mu_{ld}^{out}$, $\mu_{sv}^{out}$, and $\mu_{lv}^{out}$), as written below:
\begin{equation}
    \begin{cases}
     \mu_{ld}^{out}=f_{ld}^{dual}(\tilde{p}_{ld}~|~\gamma_{ld})
    \\
    \mu_{sv}^{out}=f_{sv}^{dual}(\tilde{p}_{sv}~|~\gamma_{sv})
     \\
    \mu_{lv}^{out}=f_{lv}^{dual}(\tilde{p}_{lv}~|~\gamma_{lv})
    \end{cases},
\label{eq4}
\end{equation}
where $\gamma_{ld}$, $\gamma_{sv}$, and $\gamma_{lv}$ are the learnable parameters.

In summary, the reconstruction subnetworks (i.e., $f_{ld}$, $f_{sv}$, and $f_{lv}$) include three modules: the PNMs (i.e., Eq. \ref{eq2}), the compensation modules (i.e., Eq. \ref{eq3}), and the DDNMs (i.e., Eq. \ref{eq4}). The network parameters $\Omega$ in Eq. \ref{eq1} are composed of learnable parameters $\beta$ and $\gamma$, i.e., $\Omega_{ld}=\{\beta_{ld}\bigcup \gamma_{ld}\}$, $\Omega_{sv}=\{\beta_{sv}\bigcup \gamma_{sv}\}$, and $\Omega_{lv}=\{\beta_{lv}\bigcup \gamma_{lv}\}$.
\subsection{Loss Function}
\label{section:C}
The overall loss function of the proposed framework consists of three parts, namely, 1) mutual learning loss among the prior images (i.e., $\mu_{ld}^{prior}$, $\mu_{sv}^{prior}$, and $\mu_{lv}^{prior}$), 2) mutual learning loss among the final outputs (i.e., $\mu_{ld}^{out}$, $\mu_{sv}^{out}$, and $\mu_{lv}^{out}$), and 3) reconstruction consistency loss between the prior images/final outputs with the low-mAs FVCT image (i.e., $\mu_{ld}$). Without loss of generality, an $\ell_{2}$ loss is adopted to construct the three above mentioned loss functions.

For simplicity, the mutual learning loss among the prior images is denoted as $\mathcal{L}_{ml}^{prior}$, as written below:
\begin{equation}
    \begin{split}
    \mathcal{L}_{ml}^{prior}=\|\mu_{ld}^{prior}-\mu_{sv}^{prior}\|_{2} + \|\mu_{ld}^{prior}-\mu_{lv}^{prior}\|_{2}
    \\
    + \|\mu_{sv}^{prior}-\mu_{lv}^{prior}\|_{2}
    \end{split},
\label{eq5}
\end{equation}
where $ml$ represents for mutual learning. Furthermore, the mutual learning loss among the final outputs is denoted as $\mathcal{L}_{ml}^{out}$:
\begin{equation}
    \begin{split}
    \mathcal{L}_{ml}^{out}=\|\mu_{ld}^{out}-\mu_{sv}^{out}\|_{2} + \|\mu_{ld}^{out}-\mu_{lv}^{out}\|_{2}
    \\
    + \|\mu_{sv}^{out}-\mu_{lv}^{out}\|_{2}
    \end{split}.
\label{eq6}
\end{equation}

In addition, the reconstruction consistency loss $\mathcal{L}_{rc}$ between the prior images/final outputs and $\mu_{ld}$ is formulated as:
\begin{equation}
    \mathcal{L}_{rc}=\|\mu_{ld}^{prior}-\mu_{ld}\|_{2} + \|\mu_{ld}^{out}-\mu_{ld}\|_{2},
\label{eq7}
\end{equation}
where $rc$ represents for reconstruction consistency. It is noted that, the reconstruction consistency losses among $\mu_{sv}^{prior}$, $\mu_{lv}^{prior}$, $\mu_{sv}^{out}$, $\mu_{lv}^{out}$ and $\mu_{ld}$ are omitted in order to reduce redundancy.

In summary, the total loss $\mathcal{L}_{total}$ for the proposed SS-CTML framework is then written below:
\begin{equation}
    \mathcal{L}_{total}=\mathcal{L}_{ml}^{prior}+\mathcal{L}_{ml}^{out}+\mathcal{L}_{rc}.
\label{eq8}
\end{equation}

With the above loss functions, the three reconstruction tasks in the SS-CTML framework can be end-to-end mutually optimized without normal-dose FVCT training data.
\section{Experimental Setup}
\subsection{Datasets}
\subsubsection{Simulated Data} The AAPM Low Dose CT Grand Challenge Dataset released by Mayo clinics (i.e., Mayo dataset) is adopted in this study. The Mayo dataset contains 10 patients with a total of 6,687 slices of normal-dose FVCT images and corresponding helical FVCT sinogram data. It should be noted that normal-dose FVCT images are only used for quantitative evaluations in our experiment. The proposed SS-CTML method utilizes eight patients data (i.e., 6,722 slices) for network training, one patient data (i.e., 525 slices) for validation and one patient data (i.e., 560 slices) for inference, respectively.
\begin{figure}[htbp]
\includegraphics[width=\columnwidth]{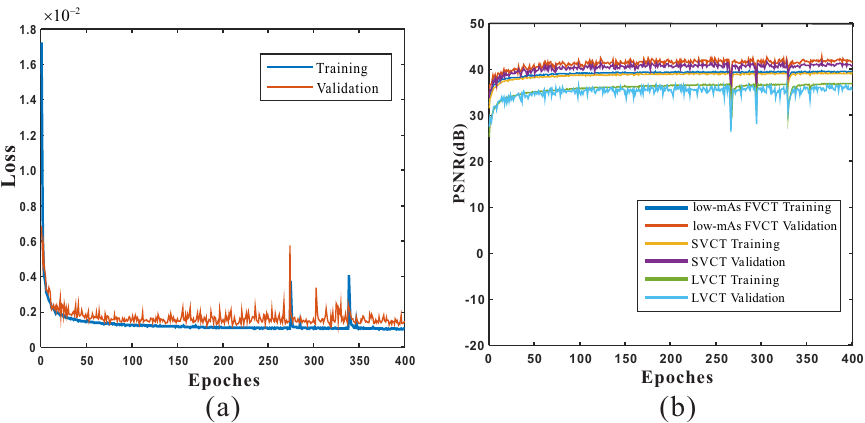}
\caption{(a) Total loss curves for training and validation; and (b) PSNR curves of the low-mAs FVCT, SVCT, and LVCT reconstruction tasks for training and validation.}
\label{fig3}
\end{figure}
\begin{figure*}[ht]
    \centering
    \centerline{\includegraphics[scale=0.94]{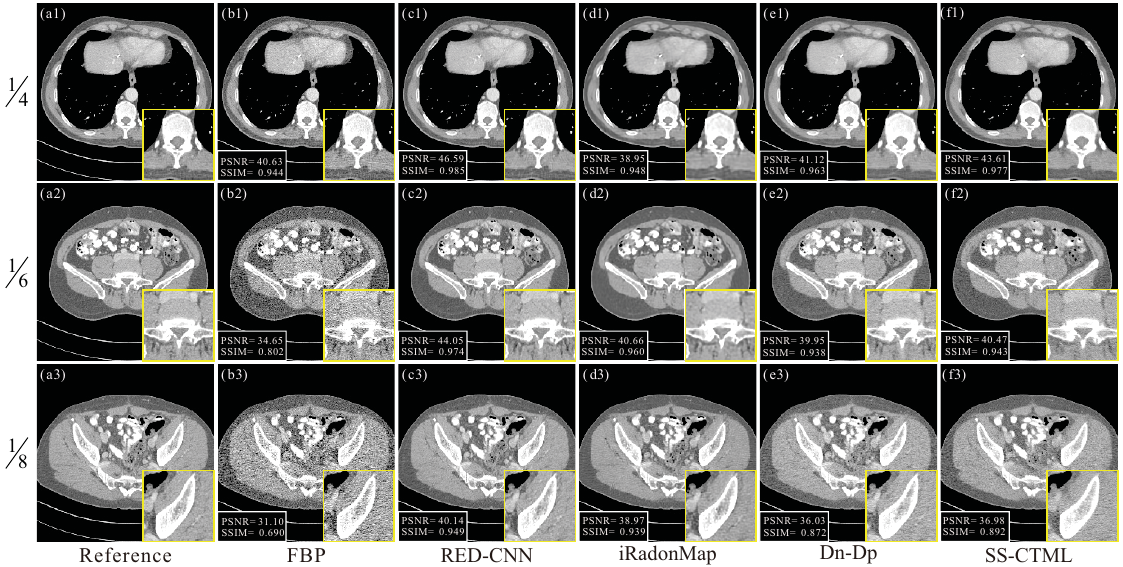}}
    \caption{The simulated data results of low-mAs FVCT reconstruction task for different methods with different dose levels. The three rows show  quarter-dose, sixth-dose, and eighth-dose results, respectively. From (a) to (f): Reference, FBP, RED-CNN, iRadonMap, Dn-DP, and SS-CTML. The display window is [-260, 340] HU.}
\label{fig4}
\end{figure*}

The corresponding helical FVCT sinogram data is first rebinned into fan-beam FVCT sinogram data with 1152 exposure views around 360 degrees. To validate the proposed framework in different low tube current/mAs situations, we simulate low-mAs FVCT sinogram data with different dose-levels, namely, quarter-dose, sixth-dose, and eighth-dose, which refers to 1/4, 1/6, and 1/8 radiation dose of the normal-dose FVCT sinogram data in the Mayo dataset, respectively. After the low-mAs FVCT sinogram data is simulated, the corresponding SVCT and LVCT sinogram data are extracted from the low-mAs FVCT sinogram data. Then, their corresponding FBP images are reconstructed. In this paper, the number of exposure view in SVCT is set to 144, which is evenly distributed around 360 degrees. The exposure view range of LVCT is set to 120 degrees, where the exposure view number is 384 within the 120 degrees.
\subsubsection{Real Clinical Data} In order to further validate the effectiveness of the proposed SS-CTML method, we test the trained SS-CTML model on real LDCT dataset. In this work, the ELCAP Public Lung Image Database, which contains 50 cases of whole-lung scanned LDCT images is adopted. The LDCT scans were obtained in a single breath hold with a 1.25mm slice thickness. Due to the lack of corresponding normal-dose CT image, quantitative index is not evaluated in the result of real clinical data.
\subsection{Implementation Details}
It is noted that various of networks can be adopted as the backbone of prior neural modules (i.e., $f_{ld}^{prior}$, $f_{sv}^{prior}$, and $f_{lv}^{prior}$) and dual-domain neural modules (i.e., $f_{ld}^{dual}$, $f_{sv}^{dual}$, and $f_{lv}^{dual}$) in the proposed SS-CTML framework. Without loss of generality, a five-stage U-net with residual learning is adopted in this study. The proposed framework was implemented with pytorch toolbox and trained on a workstation with one NVIDIA RTX 3090 graphics processing unit (GPU). Adam algorithm was employed to optimize the network parameters. Two exponential decay rates $\beta_{1}$ and $\beta_{2}$ for Adam were set to 0.5, 0.9, respectively. The initial learning rate was set to $1 \times 10^{-5}$, and the batch number was 1000. All training data are normalized to $[-1,1]$ with fixed mean and maximum value. The training and validation total loss curves and the corresponding peak signal-to-noise ratio (PSNR) curves of the low-mAs FVCT, SVCT and LVCT reconstruction tasks are shown in Fig. \ref{fig3}. As can be seen, the training process of the proposed SS-CTML method can converge quickly to a stable local minimum.
\subsection{Competing Methods}
To fully evaluate the performance of the proposed SS-CTML method, the performances of different reconstruction tasks (i.e., low-mAs FVCT, SVCT and LVCT reconstruction) are separately compared to several dedicated competing methods. In all three reconstruction tasks, the FBP reconstruction results are served as baselines. As for the low-mAs FVCT reconstruction task, we use a supervised image postprocessing method (i.e., RED-CNN\cite{chen2017low}), a supervised dual-domain learning method (i.e., iRadonMap\cite{8950464}), and a diffusion model-based method (i.e., Dn-Dp\cite{liu2023diffusion}) as the competing methods, respectively. As for the SVCT reconstruction task, a supervised image postprocessing method (i.e., FBPConvNet\cite{jin2017deep}) and two supervised dual-domain learning method (i.e., iRadonMap\cite{8950464}, and FreeSeed\cite{ma2023freeseed}) are implemented as the competing methods. As for the LVCT reconstruction task, a supervised image postprocessing method (i.e., FBPConvNet\cite{jin2017deep}), a supervised dual-domain learning method (i.e., Sam's Net\cite{chen2022sam}), and a diffusion model-based method (i.e., DOLCE\cite{liu2023dolce}) are adopted as the competing methods.
\section{Results}
\subsection{Simulated Data Results}
\begin{figure*}[ht]
    \centering
    \centerline{\includegraphics[scale=0.94]{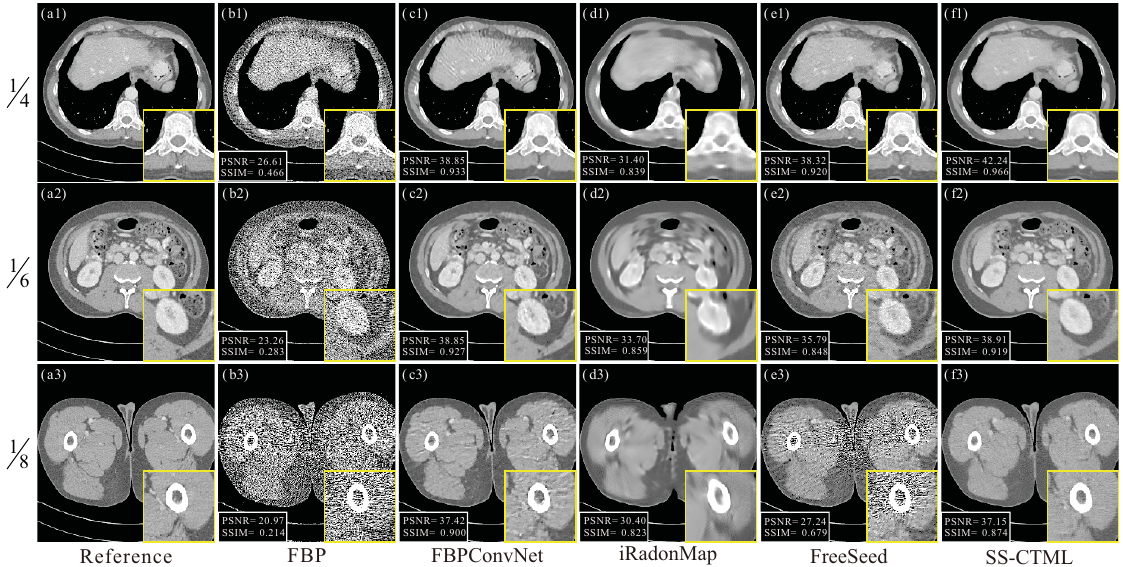}}
    \caption{The simulated data results of SVCT reconstruction task for different methods with different dose levels. The exposure view number of SVCT is set to 144. The three rows show  quarter-dose, sixth-dose, and eighth-dose results, respectively. From (a) to (f): Reference, FBP, FBPConvNet, iRadonMap, FreeSeed, and SS-CTML. The display window is [-260, 340] HU.}
\label{fig5}
\end{figure*}
\begin{figure*}[ht]
    \centering
    \centerline{\includegraphics[scale=0.94]{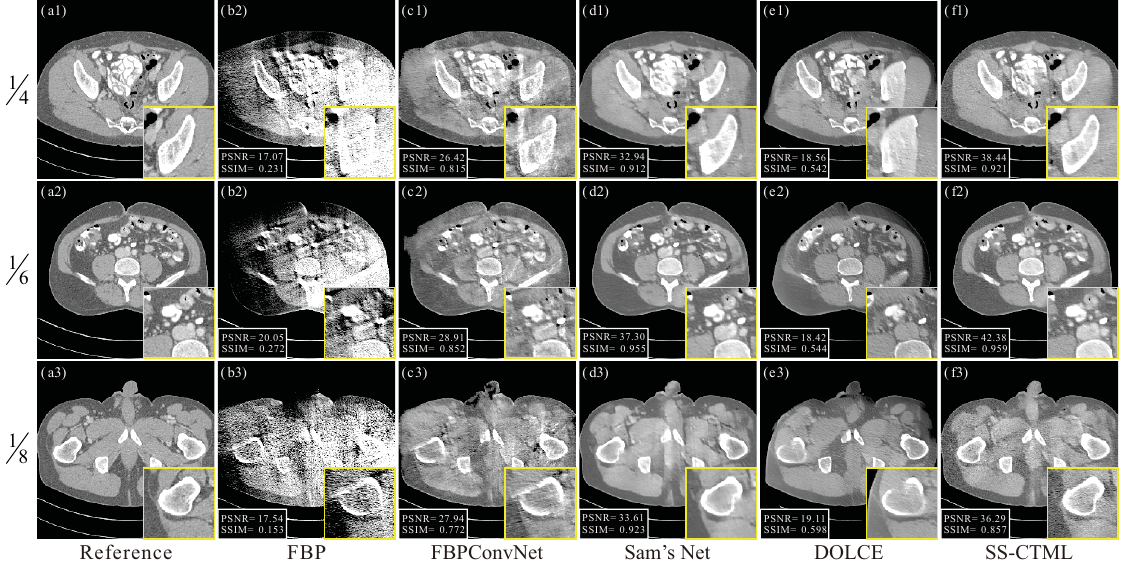}}
    \caption{The simulated data results of LVCT reconstruction task for different methods with different dose levels. The exposure view range of LVCT is set to 120 degrees. The three rows show  quarter-dose, sixth-dose, and eighth-dose results, respectively. From (a) to (f): Reference, FBP, FBPConvNet, Sam's Net, DOLCE, and SS-CTML. The display window is [-260, 340] HU.}
\label{fig6}
\end{figure*}
\subsubsection{Low-mAs FVCT Results}
\begin{figure}[b]
\centerline{\includegraphics[width=\columnwidth]{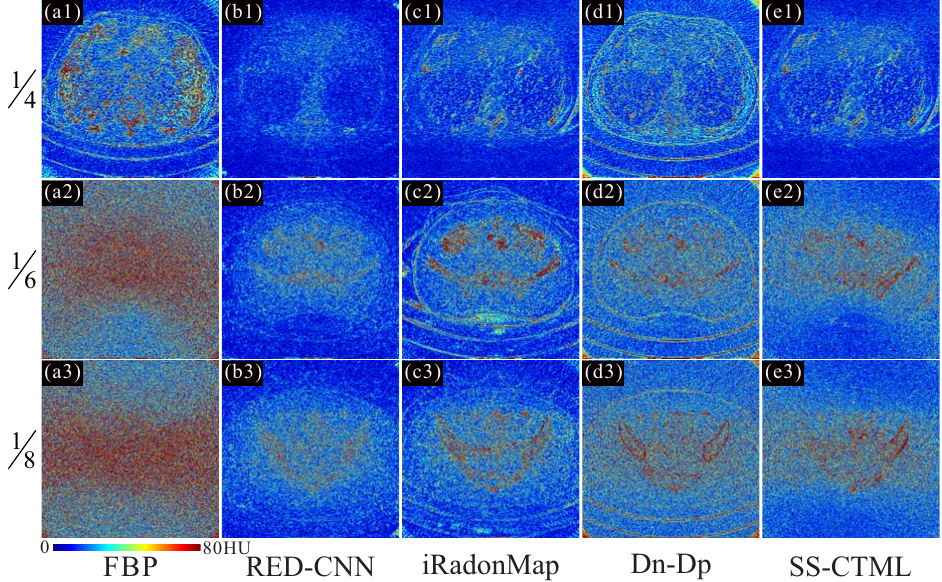}}
\caption{The absolute difference images of low-mAs FVCT reconstruction task in Fig. \ref{fig4} associated with the references. From (a) to (e): FBP, RED-CNN, iRadonMap, Dn-Dp, and SS-CTML.}
\label{fig7}
\end{figure}

The low-mAs FVCT reconstruction results are shown in Fig. \ref{fig4}. Specifically, Fig. \ref{fig4}(a) is FBP-reconstructed normal-dose FVCT image, which is served as the reference image. Fig. \ref{fig4}(b) is the results of low-mAs FVCT FBP-reconstruction. Fig. \ref{fig4}(c) and Fig. \ref{fig4}(d) show results of RED-CNN and iRadonMap. Fig. \ref{fig4}(e) is the results of Dn-Dp. Fig. \ref{fig4}(f) is the results of the proposed SS-CTML method. For intuitive comparison, a selected zoomed-in patch is shown at the lower-right corner of each image. As can be seen, the images reconstructed by FBP are contaminated with obvious noise-induced artifacts. As for the supervised-learning methods, RED-CNN can better remove noise-induced artifacts and preserve more anatomy information than iRadonMap. Although the performance of the SS-CTML method are not good as that of the supervised-learning methods mentioned above, it still has intrinsic advantages over the supervised-learning methods, since it has no need for paired training datasets. In addition, the performance of the proposed SS-CTML is also comparable to that of the diffusion model-based Dn-DP, while the inference time of our method is much shorter than that of Dn-DP. The overall quantitative evaluations of the entire testing datasets presented in Tables \ref{Tab2}, \ref{Tab3}, and \ref{Tab4} are consistent with the qualitative evaluations. Fig. \ref{fig7} presents the corresponding absolute difference images. The results of quantitative evaluations and absolute difference images are consistent with those of qualitative comparisons.
\subsubsection{SVCT Results}
\begin{figure}[tb]
\centerline{\includegraphics[width=\columnwidth]{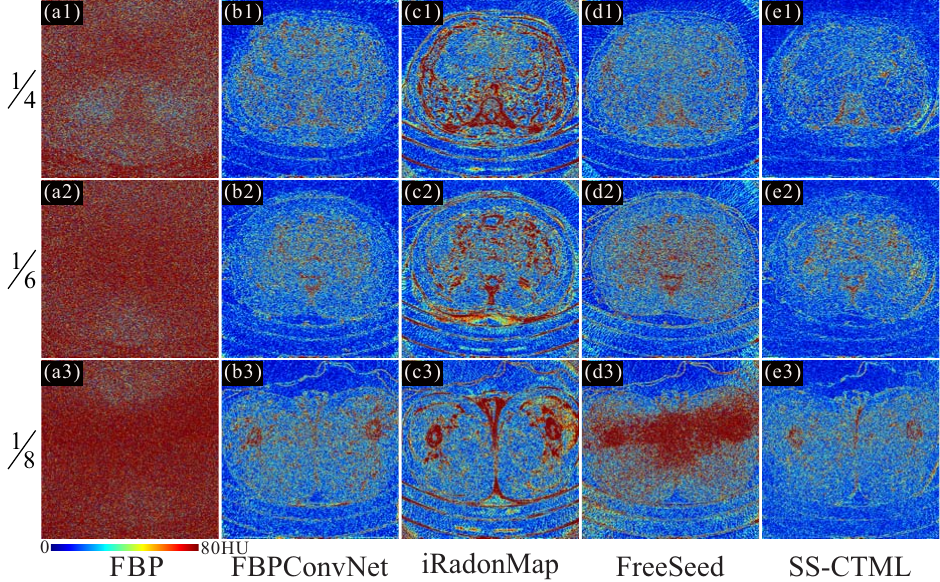}}
\caption{The absolute difference images of SVCT reconstruction task in Fig. \ref{fig5} associated with the references. From (a) to (e): FBP, FBPConvNet, iRadonMap, FreeSeed, and SS-CTML.}
\label{fig8}
\end{figure}

Fig. \ref{fig5} shows the results of SVCT reconstruction under different dose-levels. The layout of Fig. \ref{fig5} is same as that of Fig. \ref{fig4}. The Fig. \ref{fig5}(a) and \ref{fig5}(b) show the normal-dose FBP-reconstructed results and low-mAs SVCT FBP-reconstructed results, respectively. Fig. \ref{fig5}(c) and \ref{fig5}(d) are results of two supervised-learning methods in image domain and dual-domain, i.e., FBPConvNet and iRadonMap. Fig. \ref{fig5}(e) is results of FreeSeed, which removes streak artifacts of SVCT images in fourier domain. It is observed that due to the irregularity overlay of streak artifacts and noises, the restoration abilities of FBPConvNet and iRadonMap are limited. Although FreeSeed method can restore the SVCT images effectively under quarter-dose condition, the restoration performance decreases when streak artifacts increase under other conditions. The streak artifacts are still existing under sixth-dose and eighth-dose conditions, as can be observed in Fig. \ref{fig5}(e2) and Fig. \ref{fig5}(e3). The qualitative comparison in Fig. \ref{fig5} and quantitative evaluations in Tables \ref{Tab2}, \ref{Tab3}, and \ref{Tab4} indicate that the proposed SS-CTML method can better remove noise and streak artifacts than the competing supervised learning methods in SVCT reconstruction task. Fig. \ref{fig8} shows the corresponding absolute difference images. We can observe that the proposed SS-CTML can maintain stable SVCT image restoration performance under different dose conditions.
\begin{table}[t]
    \caption{Quantitative measures of the three CT reconstruction tasks for different methods with quarter-dose data.}
    \centering
    \renewcommand{\arraystretch}{1.3}
    \begin{tabular}{p{5 mm}<{\raggedright}p{13 mm}<{\raggedright}p{15 mm}<{\centering}p{16 mm}<{\centering}p{16 mm}<{\centering}}
    \toprule
    \multirow{2}*{Tasks}& \multirow{2}*{Methods}& \multicolumn{3}{c}{Quarter-Dose} \\
    \cline{3-5}
     ~&~& PSNR & NMSE & SSIM  \\
    \hline
    \multirow{5}*{FVCT}
    &FBP           &34.34~$\pm$~3.5630& 0.0033~$\pm$~0.0001& 0.800~$\pm$~0.0002 \\
    ~&RED-CNN      &\textbf{41.78~$\pm$~2.4488}& \textbf{0.0006~$\pm$~0.0001}& \textbf{0.956~$\pm$~0.0002}\\
    ~&iRadonMap    &38.15~$\pm$~0.9711& 0.0017~$\pm$~0.0001& 0.929~$\pm$~0.0001\\
    ~&Dn-Dp        &38.12~$\pm$~1.4099& 0.0014~$\pm$~0.0001& 0.917~$\pm$~0.0005\\
    ~&SS-CTML      &39.58~$\pm$~2.1694& 0.0010~$\pm$~0.0001& 0.932~$\pm$~0.0004\\
    \hline
    \multirow{5}*{SVCT}
    &FBP           &24.09~$\pm$~2.0305& 0.0331~$\pm$~0.0001& 0.332~$\pm$~0.0039 \\
    ~&FBPConvNet   &37.89~$\pm$~1.7773& 0.0018~$\pm$~0.0001& 0.917~$\pm$~0.0004\\
    ~&iRadonMap    &31.44~$\pm$~1.0609& 0.0125~$\pm$~0.0001& 0.839~$\pm$~0.0004\\
    ~&FreeSeed     &36.12~$\pm$~2.0959& 0.0028~$\pm$~0.0001& 0.872~$\pm$~0.0012\\
    ~&SS-CTML      &\textbf{39.25~$\pm$~1.2294}& \textbf{0.0012~$\pm$~0.0001}& \textbf{0.939~$\pm$~0.0002}\\
    \hline
    \multirow{5}*{LVCT}
    &FBP           &17.46~$\pm$~0.7784& 0.1625~$\pm$~0.0003& 0.246~$\pm$~0.0015 \\
    ~&FBPConvNet   &26.16~$\pm$~3.7332& 0.0876~$\pm$~0.0007& 0.819~$\pm$~0.0008\\
    ~&Sam's Net    &32.56~$\pm$~5.0788& 0.0153~$\pm$~0.0002& \textbf{0.910~$\pm$~0.0003}\\
    ~&DOLCE        &18.84~$\pm$~1.7557& 0.1299~$\pm$~0.0033& 0.552~$\pm$~0.0067\\
    ~&SS-CTML      &\textbf{33.92~$\pm$~3.2714}& \textbf{0.0080~$\pm$~0.0001}& {0.909~$\pm$~0.0004}\\
    \bottomrule
    \end{tabular}
    \label{Tab2}
\end{table}
\begin{table}[t]
    \caption{Quantitative measures of the three CT reconstruction tasks for different methods with sixth-dose data.}
    \centering
    \renewcommand{\arraystretch}{1.3}
    \begin{tabular}{p{5 mm}<{\raggedright}p{13 mm}<{\raggedright}p{15 mm}<{\centering}p{16 mm}<{\centering}p{16 mm}<{\centering}}
    \toprule
    \multirow{2}*{Tasks}& \multirow{2}*{Methods}& \multicolumn{3}{c}{Sixth-Dose} \\
    \cline{3-5}
     ~&~& PSNR & NMSE & SSIM  \\
    \hline
    \multirow{5}*{FVCT}
    &FBP           &32.25~$\pm$~3.7528& 0.0054~$\pm$~0.0001& 0.722~$\pm$~0.0052 \\
    ~&RED-CNN      &\textbf{41.07~$\pm$~2.3933}& \textbf{0.0008~$\pm$~0.0001}& \textbf{0.949~$\pm$~0.0002}\\
    ~&iRadonMap    &39.40~$\pm$~1.1446& 0.0012~$\pm$~0.0001& 0.939~$\pm$~0.0002\\
    ~&Dn-Dp        &37.13~$\pm$~1.9409& 0.0017~$\pm$~0.0001& 0.893~$\pm$~0.0011\\
    ~&SS-CTML      &38.08~$\pm$~2.6019& 0.0014~$\pm$~0.0001& 0.904~$\pm$~0.0009\\
    \hline
    \multirow{5}*{SVCT}
    &FBP           &22.41~$\pm$~2.5229& 0.0492~$\pm$~0.0001& 0.271~$\pm$~0.0038 \\
    ~&FBPConvNet   &37.70~$\pm$~1.6234& 0.0018~$\pm$~0.0001& 0.914~$\pm$~0.0004\\
    ~&iRadonMap    &31.97~$\pm$~1.2078& 0.0113~$\pm$~0.0001& 0.857~$\pm$~0.0004\\
    ~&FreeSeed     &33.19~$\pm$~6.4637& 0.0078~$\pm$~0.0001& 0.803~$\pm$~0.0044\\
    ~&SS-CTML      &\textbf{38.52~$\pm$~1.2810}& \textbf{0.0014~$\pm$~0.0001}& \textbf{0.929~$\pm$~0.0002}\\
    \hline
    \multirow{5}*{LVCT}
    &FBP           &17.26~$\pm$~0.7893& 0.1683~$\pm$~0.0002& 0.203~$\pm$~0.0017\\
    ~&FBPConvNet   &26.09~$\pm$~3.1627& 0.0797~$\pm$~0.0007& 0.778~$\pm$~0.0012\\
    ~&Sam's Net    &33.72~$\pm$~2.7168& 0.0089~$\pm$~0.0001& \textbf{0.922~$\pm$~0.0003}\\
    ~&DOLCE        &18.29~$\pm$~2.4573& 0.1353~$\pm$~0.0027& 0.565~$\pm$~0.0082\\
    ~&SS-CTML      &\textbf{33.86~$\pm$~3.0895}& \textbf{0.0080~$\pm$~0.0001}& 0.894~$\pm$~0.0006\\
    \bottomrule
    \end{tabular}
    \label{Tab3}
\end{table}
\begin{table}[t]
    \caption{Quantitative measures of the three CT reconstruction tasks for different methods with eighth-dose data.}
    \centering
    \renewcommand{\arraystretch}{1.3}
    \begin{tabular}{p{5 mm}<{\raggedright}p{13 mm}<{\raggedright}p{15 mm}<{\centering}p{16 mm}<{\centering}p{16 mm}<{\centering}}
    \toprule
    \multirow{2}*{Tasks}& \multirow{2}*{Methods}& \multicolumn{3}{c}{Eighth-Dose} \\
    \cline{3-5}
     ~&~& PSNR & NMSE & SSIM  \\
    \hline
    \multirow{5}*{FVCT}
    &FBP           &30.82~$\pm$~3.8350& 0.0076~$\pm$~0.0001& 0.661~$\pm$~0.0066 \\
    ~&RED-CNN      &\textbf{40.51~$\pm$~2.4985}& \textbf{0.0009~$\pm$~0.0001}& \textbf{0.944~$\pm$~0.0003}\\
    ~&iRadonMap    &38.00~$\pm$~1.0810& 0.0018~$\pm$~0.0001& 0.933~$\pm$~0.0001\\
    ~&Dn-Dp        &36.31~$\pm$~2.5227& 0.0021~$\pm$~0.0001& 0.870~$\pm$~0.0019\\
    ~&SS-CTML      &36.96~$\pm$~2.7884& 0.0018~$\pm$~0.0001& 0.878~$\pm$~0.0014\\
    \hline
    \multirow{5}*{SVCT}
    &FBP           &21.19~$\pm$~2.8197& 0.0660~$\pm$~0.0002& 0.234~$\pm$~0.0035 \\
    ~&FBPConvNet   &36.72~$\pm$~1.8194& 0.0024~$\pm$~0.0001& 0.897~$\pm$~0.0007\\
    ~&iRadonMap    &30.79~$\pm$~0.9910& 0.0146~$\pm$~0.0001& 0.833~$\pm$~0.0004\\
    ~&FreeSeed     &30.07~$\pm$~13.016& 0.0197~$\pm$~0.0002& 0.723~$\pm$~0.0097\\
    ~&SS-CTML      &\textbf{37.83~$\pm$~1.3742}& \textbf{0.0017~$\pm$~0.0001}& \textbf{0.918~$\pm$~0.0004}\\
    \hline
    \multirow{5}*{LVCT}
    &FBP           &17.07~$\pm$~0.8097& 0.1743~$\pm$~0.0002& 0.175~$\pm$~0.0018 \\
    ~&FBPConvNet   &25.78~$\pm$~3.1906& 0.0874~$\pm$~0.0010& 0.773~$\pm$~0.0019\\
    ~&Sam's Net    &32.57~$\pm$~3.4341& 0.0133~$\pm$~0.0001& \textbf{0.912~$\pm$~0.0003}\\
    ~&DOLCE        &19.03~$\pm$~1.0382& 0.1563~$\pm$~0.0053& 0.591~$\pm$~0.0058\\
    ~&SS-CTML      &\textbf{33.34~$\pm$~2.4266}& \textbf{0.0080~$\pm$~0.0001}& 0.873~$\pm$~0.0009\\
    \bottomrule
    \end{tabular}
    \label{Tab4}
\end{table}
\begin{figure}[tbp]
\centerline{\includegraphics[width=\columnwidth]{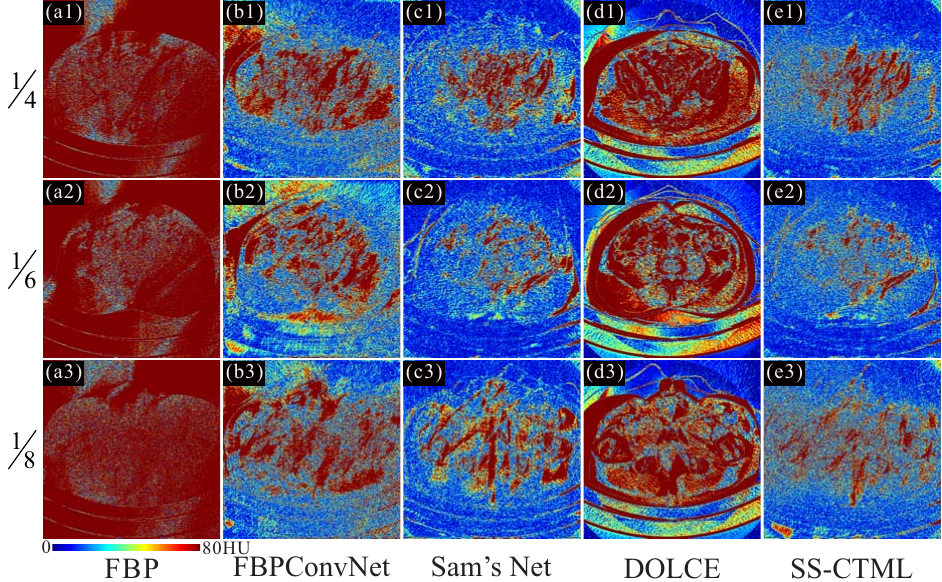}}
\caption{The absolute difference images of SVCT reconstruction task in Fig. \ref{fig6} associated with the references. From (a) to (e): FBP, FBPConvNet, Sam's Net, DOLCE, and SS-CTML.}
\label{fig9}
\end{figure}
\subsubsection{LVCT Results}
Fig. \ref{fig6} presents the LVCT reconstruction results with different dose-levels. From Fig. \ref{fig6}(a) to Fig. \ref{fig6}(f) are the reference image, FBP reconstructed results of simulated limited-view data, FBPConvNet, Sam's Net, DOLCE and the proposed SS-CTML method. Due to the absence of some specific exposure angles, the anatomy structures in Fig. \ref{fig6}(b) are distorted. In the supervised network reconstruction results, Sam's Net method can better restore distorted anatomy structures than FBPConvNet. As a diffusion model-based LVCT reconstruction method, DOLCE is unable to generate promising reconstruction performance in this study, probably due to the complicated characteristics of the constructed LVCT testing dataset, where both limited-view and low-mAs artifacts are mixed together. By analysing the results in Fig. \ref{fig6} and quantitative comparisons in Tables \ref{Tab2}, \ref{Tab3}, and \ref{Tab4}, we can find that the SS-CTML method can better restore the distorted anatomy structure and low contrast soft tissue information in LVCT images than the competing supervised and unsupervised methods. Fig. \ref{fig9} exhibits the corresponding absolute difference images. It can be concluded that the proposed SS-CTML method achieves the lower difference numerical value than other methods.
\subsection{Real Data Results}
\begin{figure}[tb]
\centerline{\includegraphics[width=\columnwidth]{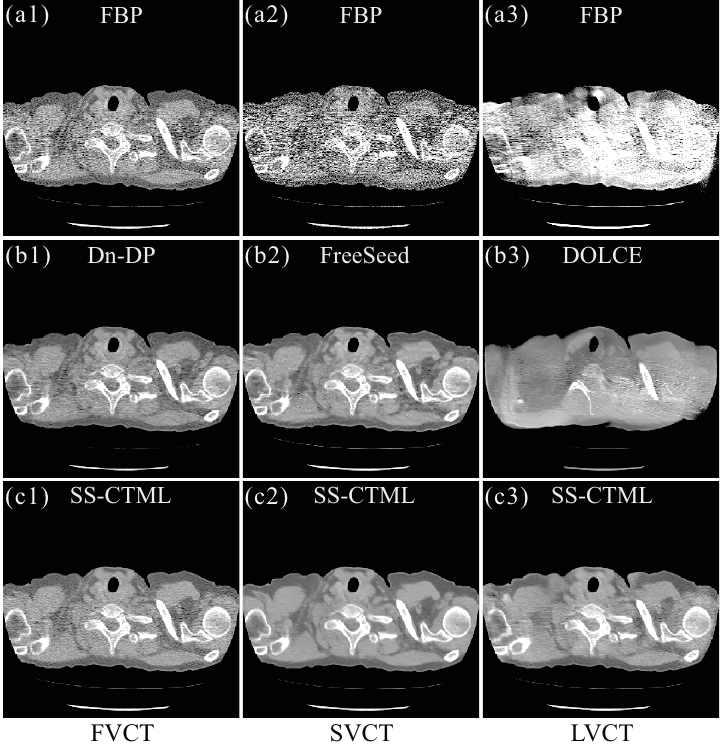}}
\caption{The real clinical data results of different CT reconstruction tasks. From first to last column: FVCT, SVCT, LVCT reconstruction task, respectively. The display windows is [-260, 340] HU.}
\label{fig10}
\end{figure}

The reconstruction results of real clinical data are displayed in Fig. \ref{fig10}. It is mentioned that all the reconstruction results in Fig. \ref{fig10} are generated by the competing methods trained with quarter-dose Mayo data. The three columns show the low-mAs FVCT, SVCT and LVCT reconstruction results, respectively. It can be observed that the SS-CTML method can achieve better performance than the supervised and unsupervised methods, especially in SVCT and LVCT reconstruction tasks.

\subsection{Ablation Study}
In this section, we conduct several indispensable ablation experiments to evaluate the effectiveness of different components in SS-CTML framework. Specifically, to illustrate the importance of PNM and DDNM, we conduct ablation experiments by excluding one of them to measure the reconstruction performance. In addition, to validate the effectiveness of multi-task learning scheme, the mutual learning of two reconstruction tasks is also conducted. The quantitative results are listed in Table \ref{Tab5}. Specifically, w/o PNM means the prior neural module is not included in the SS-CTML optimization, while w/o DDNM means the dual-domain neural module is not included in the SS-CTML optimization. The mutual learning of two reconstruction tasks contains three forms, which is that without FVCT reconstruction task, without SVCT reconstruction task and without LVCT reconstruction task, i.e. w/o FVCT, w/o SVCT and w/o LVCT, respectively. As we can see, in Table \ref{Tab5}, the performances of SS-CTML would decrease without either the PNM or DDNM. It is further observed that the multi-task mutual learning strategy makes great contribution to the final results, by comparing the reconstruction performances of the two-task and three-task strategies.

\begin{table}[hb!]
    \caption{Ablation study on critical components in the SS-CTML framework.}
    \centering
    \renewcommand{\arraystretch}{1.3}
    \begin{tabular}{p{11 mm}<{\raggedright}p{7 mm}<{\raggedright}p{15 mm}<{\centering}p{16 mm}<{\centering}p{16 mm}<{\centering}}
    \toprule
    \multirow{2}*{Variants}& \multirow{2}*{Tasks}& \multicolumn{3}{c}{Quarter-Dose}  \\
    \cline{3-5}
     ~&~& PSNR & NMSE & SSIM \\
    \hline
    \multirow{3}*{SS-CTML} &FVCT  &\textbf{39.58~$\pm$~2.1694}& \textbf{0.0010~$\pm$~0.0001} &0.932~$\pm$~0.0004\\
    ~&SVCT                        &39.25~$\pm$~1.2294& 0.0012~$\pm$~0.0001 &\textbf{0.939~$\pm$~0.0002}\\
    ~&LVCT                        &\textbf{33.92~$\pm$~3.2714}& \textbf{0.0080~$\pm$~0.0001} &\textbf{0.909~$\pm$~0.0004}\\
    \hline
    \multirow{3}*{w/o PNM} &FVCT  &38.81~$\pm$~1.6248& 0.0014~$\pm$~0.0001 & \textbf{0.933~$\pm$~0.0004}\\
    ~&SVCT                        &38.22~$\pm$~1.1419& 0.0016~$\pm$~0.0001 &0.933~$\pm$~0.0002\\
    ~&LVCT                        &32.32~$\pm$~2.3657& 0.0137~$\pm$~0.0002 &0.900~$\pm$~0.0002\\
    \hline
    \multirow{3}*{w/o DDNM}&FVCT  &39.02~$\pm$~1.9401& 0.0012~$\pm$~0.0001 & 0.931~$\pm$~0.0004 \\
    ~&SVCT                        &38.55~$\pm$~1.2628& 0.0015~$\pm$~0.0001 & 0.934~$\pm$~0.0002 \\
    ~&LVCT                        &32.10~$\pm$~2.6495& 0.0194~$\pm$~0.0002 & 0.901~$\pm$~0.0005 \\
    \hline
    \multirow{2}*{w/o FVCT}
    &SVCT     &29.10~$\pm$~1.5929& 0.0111~$\pm$~0.0001 & 0.589~$\pm$~0.0003 \\
    ~&LVCT    &32.66~$\pm$~2.0045& 0.0084~$\pm$~0.0001 & 0.850~$\pm$~0.0004 \\
    \hline
    \multirow{2}*{w/o SVCT}
    &FVCT        &37.37~$\pm$~2.9096& 0.0017~$\pm$~0.0001 & 0.893~$\pm$~0.0012 \\
    ~&LVCT   &32.93~$\pm$~1.9436& 0.0094~$\pm$~0.0001 & 0.882~$\pm$~0.0005 \\
     \hline
    \multirow{2}*{w/o LVCT}
     &FVCT         &39.19~$\pm$~3.3678& 0.0010~$\pm$~0.0001 & 0.915~$\pm$~0.0008 \\
    ~&SVCT         &\textbf{39.34~$\pm$~1.3795}&\textbf{0.0011~$\pm$~0.0001} & 0.931~$\pm$~0.0003 \\
    \bottomrule
    \end{tabular}
    \label{Tab5}
\end{table}

\section{Discussion}
The current work has some limitations. The first limitation is backbone network. The backbone of the proposed SS-CTML framework is a five-stage U-net with residual learning, which has not been specifically designed in this study. In future work, more advanced backbone network (such as, Transformer and/or Mamba) would be considered to improve the reconstruction performance. The second limitation is that the SS-CTML framework needs three subnetworks for the three different tasks, which is still a redundant architecture. In the future, we will try to adopt one single network to handle the three tasks to reduce redundancy. Lastly, the current SS-CTML framework is trained under condition of a fixed dose-level, fixed sparse-view sampling rate and fixed limited-view range, which requires retraining our model to adapt data with different imaging settings. Therefore, in future work, we will attempt to train networks with different dose-level and imaging geometries simultaneously to meet the needs of real clinical applications. It is noted that, by solving the above limitations, the SS-CTML framework has great potential to serve as a foundation model for CT image reconstruction, only if we have more computing resources to enlarge the network capacity and to include more CT imaging tasks (such as, metal artifact reduction, MAR). Nevertheless, it is still a prospective yet challenging endeavor to develop a CT image reconstruction foundation model based on the presented framework, and we will contribute more in future work.
\section{Conclusion}
In this paper, we propose a self-supervised cross-task mutual learning framework (i.e., SS-CTML) for CT image reconstruction, which can handle low-mAs FVCT, SVCT and LVCT reconstruction tasks independently. The SS-CTML framework mainly consists of prior neural modules (PNMs) and dual-domain neural modules (DDNMs). We validate the the effectiveness of the proposed SS-CTML framework with both simulated and real LDCT data. The experimental results demonstrate that the proposed method can achieve promising CT image reconstruction performance with both quantitative and qualitative measurements.

\end{document}